\documentstyle[12pt,aaspp4,psfig]{article}
\def\pp{\par\parshape 2 0truecm 15.5truecm 1truecm 14.5truecm\noindent}
\newcommand{\simgt}{\lower.5ex\hbox{$\; \buildrel > \over \sim \;$}}
\newcommand{\simlt}{\lower.5ex\hbox{$\; \buildrel < \over \sim \;$}}

\newcommand{\dm}{{\rm\scriptscriptstyle DM}}
\begin{document}

\begin{minipage}[c]{4cm}
RESCEU-36/98\\
UTAP-296/98\\
ADAC-030\\
astro-ph/9810247
\end{minipage}\\

\title{{\Large Reconstructing the radial profiles} \\
{\Large of gas density and temperature in clusters of galaxies} \\
{\Large from high-resolution X-ray and radio observations} 
}

\bigskip

\author{Kohji Yoshikawa \\ {\it Department of Astronomy, Kyoto
University, Kyoto 606-8502, Japan.} \\ and \\ 
  Yasushi Suto \\ {\it Department of Physics and Research Center for
    the Early Universe (RESCEU) \\ School of Science, University of
    Tokyo, Tokyo 113-0033, Japan.} }

\bigskip

\affil{ e-mail: kohji@kusastro.kyoto-u.ac.jp, 
suto@phys.s.u-tokyo.ac.jp}

\bigskip

\received{1998 May 26}
\accepted{1998 October 15}

\begin{abstract}
  We describe a non-parametric method to reconstruct gas density and
  temperature profiles of galaxy clusters from observations of X-ray
  surface brightness, emission-weighted temperature and the
  $y$-parameter through the Sunyaev-Zel'dovich effect.  This method is
  based on the inversion of the projected profiles under the
  spherically symmetric approximation, which is independent of the
  equation of state of cluster gas and does not assume that the gas is
  in hydrostatic equilibrium.  In particular we examine the
  reliability of the reconstruction method assuming a few theoretical
  models for cluster gas and assigning the statistical errors expected
  for future X-ray and radio observations.  We also discuss briefly
  the effect of non-sphericity on the basis of simulated clusters.
\end{abstract}

\keywords{ cosmology: theory -- dark matter -- galaxies: clusters:
  general -- X-rays: galaxies }

\vfill

\centerline{\it The Astrophysical Journal Part 1, in press}

\clearpage

\baselineskip14pt
\parskip2pt

\section{Introduction}

The importance of cluster abundances and baryon fractions is
well-recognized and they are now regarded as one of the standard
established statistics to constrain the cosmological models (e.g.,
Henry \& Arnaud 1991; White, Efstathiou, \& Frenk 1993; White et al.
1993; Viana \& Liddle 1996; Eke, Cole, \& Frenk 1996; Kitayama \& Suto
1997; Kitayama, Sasaki \& Suto 1998; Shimasaku 1998). On observational
sides, the accuracy and reliability of such statistics will be
considerably improved with several upcoming projects. In particular
the angular resolutions of the AXAF (Advanced X-ray Astrophysics
Facility) and XMM (X-Ray Multi-Mirror Mission) are $0.5''$ in 0.1-10
keV and $15''$ in 0.1-15 keV, respectively, and therefore they are
expected to provide high-resolution X-ray surface brightness and
temperature maps for many clusters with unprecedented angular and
energy resolutions. The existing interferometric facilities in radio
bands including BIMA (Berkeley Illinois Maryland Association) and OVRO
(Owens Valley Radio Observatory) (e.g., Cooray et al. 1998) also
produce two-dimensional maps of clusters via the Sunyaev-Zel'dovich
(SZ) effect with $10''\sim 30''$ resolution and signal-to-noise ratio
more than 20. Furthermore, the Japanese Large Millimeter and
Submillimeter Array ({\it LMSA}) project, for instance, will start an
extensive survey of clusters in millimeter and submillimeter bands
(see also Komatsu et al. 1998).  Such detailed information of
intracluster gas in multi-bands will significantly improve the role of
clusters as cosmological probes.

On theoretical sides, recent N-body/hydrodynamical simulations (e.g.,
Couchman et al. 1995; Bryan \& Norman 1998; Eke \& Frenk 1998;
Yoshikawa, Itoh, \& Suto 1998) start to reveal several important keys
to understanding the intracluster gas on a physical basis. In
particular, Navarro, Frenk \& White (1997; NFW hereafter) suggest that
a dark matter halo of cluster is described by a universal density
profile, or at least by a class of well-specified functional forms
(see Fukushige \& Makino 1997; Moore et al. 1998). Their results imply
that with appropriate equation of state, the gas density and
temperature profiles of clusters can be predicted given a cosmological
model, at least in a spherically symmetric approximation (Makino,
Sasaki \& Suto 1998; Suto, Sasaki, \& Makino 1998). In other words,
one does not have to adopt too simplified empirical models like the
isothermal $\beta$-model in approximating the cluster gas. In fact we
are now in a position to construct a more physical model combining the
new observational data and the theoretical predictions. By
reconstructing the cluster profiles in a model-independent manner from
the observations and comparing them with the theoretical models, one
can directly test the universal dark halo conjecture, and if
confirmed, probe the cosmological parameters in an independent and
complementary manner with more conventional approaches like cluster
abundances, cosmic microwave background anisotropies, and galaxy
clustering statistics (e.g., Tegmark et al. 1998). Therefore a
reliable and model-independent reconstruction of radial profiles of
cluster gas should also contribute significantly to the conventional
analyses of clusters including the estimates of the Hubble constant
(Silk \& White 1978; Inagaki, Suginohara, \& Suto 1995; Kobayashi,
Sasaki \& Suto 1996) and peculiar velocity (e.g., Yoshikawa, Itoh, \&
Suto 1998), baryon fraction and gravitational lensing (e.g., Wu \&
Fang 1997). See Birkinshaw (1998) for an excellent review on the SZ
effect and its cosmological implications.

In this paper, we apply an idea proposed earlier by Silk \& White
(1978), and examine in detail a non-parametric reconstruction method
of the radial profiles of clusters, which does not depend on any
assumption such as hydrostatic equilibrium and equation of state of
cluster gas except for the spherical symmetry of clusters. We also
propose a new method to reconstruct the temperature radial profile of
clusters only from X-ray observations. This method is also applicable
to other elliptic models as well, provided that the ellipticity is
assumed a priori (c.f., Fabricant, Rybicki \& Gorenstein 1984; Hughes
\& Birkinshaw 1998).  In this context, we note that Zaroubi et al.
(1998) recently proposed a different reconstruction method of
3-dimensional profiles of clusters using the Fourier Slice Theorem,
and our current scientific motivation is quite similar with theirs.
While their method, employing Fourier transform of the projected
images and inverse transformation, is fairly complicated and it is not
clear yet to what extent it is practical, our procedure is rather
straightforward and useful.

\section{Abel's integral solution for radial profiles of
gas density and temperature in clusters of galaxies}

Most observable quantities for a cluster are often written as an
integration over the line-of-sight:
\begin{equation} \label{eq:ftheta} f(\theta) = \int_{-\infty}^\infty
g(r) dl = 2 \int_{d_A\theta}^\infty g(r) {r dr \over \sqrt{r^2-
d_A^2\theta^2}} , \end{equation}
where the cluster is assumed to be spherically symmetric, $\theta$ and
$r$ denote the (projected) angular separation and (3D) spatial radius
from the cluster center, and $d_A$ is the angular diameter distance to
the cluster. Using Abel's integral, equation (\ref{eq:ftheta}) can be
inverted to give
\begin{equation} \label{eq:gr} g(r) = {1 \over \pi d_A}
\int_{\infty}^{r/d_A} {d f(\theta) \over d\theta} { d\theta \over
\sqrt{\theta^2 - r^2/d_A^2}} . \end{equation}

The above Abel's integral solution can be readily applied to the
bolometric X-ray surface brightness:
\begin{eqnarray} \label{eq:sx}
  S_x(\theta) 
= A_x \int_{-\infty}^{\infty} \alpha_{\rm X}(T_{\rm e}) n_{\rm
  e}^2(\sqrt{\theta^2 d_A^2(z)+l^2})~dl
= 2 A_x \int_{d_A\theta}^\infty \alpha_{\rm X}(T_{\rm e})
  n_e^2(r) {r dr \over \sqrt{r^2- d_A^2\theta^2}}, 
\end{eqnarray}
and y-parameter, or more generally, the SZ flux at a given frequency
band:
\begin{eqnarray}
\label{eq:sy}
  S_y(\theta) = 2 A_y \int_{d_A\theta}^\infty T_e(r)
  n_e(r) {r dr \over \sqrt{r^2- d_A^2\theta^2}}.
\end{eqnarray}
In the above expressions, $T_e(r)$ and $n_e(r)$ are the temperature
and number density of the electron gas of the cluster, and
$\alpha_{\rm X}(T_{\rm e})$ is the X-ray emissivity. While the
coefficients $A_x$ and $A_y$ can be explicitly computed once the
observational bands are specified, their specific values are not
important in the following analysis. Silk \& White (1978) were the
first to propose to apply the Abel inversion to the X-ray and SZ
profiles, but their discussion is limited to the estimation of the
central values of gas density and temperature aiming at the
determination of $d_A(z)$.  The purpose of the present paper is to
examine the feasibility of the reconstruction method quantitatively,
properly taking account of the current and upcoming data quality.

If one considers only the thermal bremsstrahlung for the X-ray
emissivity, which is a good approximation for clusters with $T_e
\simgt 3$keV, then $\alpha_{\rm X}(T_{\rm e})=T_{\rm e}^{1/2}(r)$, and
equations (\ref{eq:sx}) and (\ref{eq:sy}) are inverted to yield
\begin{equation}
  [n_e(r)]^3 \, = \,{A_y \over \pi A_x^2 d_A} 
  {\displaystyle \left[ \int_{\infty}^{r/d_A} 
    {d S_x(\theta) \over d\theta}
    { d\theta \over \sqrt{\theta^2 - r^2/d_A^2}} \right]^2
  \over
  \displaystyle  \int_{\infty}^{r/d_A} 
  {d S_y(\theta) \over d\theta}
  { d\theta \over \sqrt{\theta^2 - r^2/d_A^2}} 
  }, \\
\label{eq:nprof}
\end{equation}
\begin{equation}
 \left[T_e(r)\right]^2/ \alpha(T_{\rm e})\, = \, 
  \left[T_e(r)\right]^{3/2} \, = \,
  {A_x \over \pi A_y^2 d_A} 
  {\displaystyle \left[ \int_{\infty}^{r/d_A} 
    {d S_y(\theta) \over d\theta}
    { d\theta \over \sqrt{\theta^2 - r^2/d_A^2}} \right]^2
  \over
  \displaystyle  \int_{\infty}^{r/d_A} 
  {d S_x(\theta) \over d\theta}
  { d\theta \over \sqrt{\theta^2 - r^2/d_A^2}} 
  }.
  \label{eq:tprof1} 
\end{equation}
In reality, the observed X-ray flux is band-limited and one should also
take into account other emission processes than the thermal
bremsstrahlung. While the present methodology works for those cases,
the result cannot be expressed in a simple form as the above equation,
and we focus on the case of $\alpha_{\rm X}(T_{\rm e})=T_{\rm
e}^{1/2}(r)$ as an explicit example.

If the emission-weighted temperature profile projected on the sky:
\begin{equation}
  T_{\rm X}(\theta) \equiv 
{
{\displaystyle 
\int_{-\infty}^{\infty} T_{\rm e}\alpha_{\rm X}(T_{\rm e}) n_{\rm
  e}^2(\sqrt{d_A^2\theta^2 +l^2})~dl }
   \over 
{\displaystyle \int_{-\infty}^{\infty} \alpha_{\rm X}(T_{\rm e}) n_{\rm
  e}^2(\sqrt{d_A^2\theta^2+l^2})~dl }
} 
= {
{\displaystyle 
\int_{-\infty}^{\infty} T_{\rm e}\alpha_{\rm X}(T_{\rm e}) n_{\rm
  e}^2(\sqrt{d_A^2\theta^2+l^2})~dl }
   \over 
S_x(\theta)/A_x } ,
\label{eq:txproj}
\end{equation}
is measured, however, one can similarly derive
\begin{eqnarray}
 T_{\rm e}(r)\, \alpha_{\rm X}(T_{\rm e}) \, [n_e(r)]^2 
 &=& {1 \over \pi A_x d_A} 
\int_{\infty}^{r/d_A} 
{d [S_x(\theta)T_{\rm X}(\theta)] \over d\theta}
{ d\theta \over \sqrt{\theta^2 - r^2/d_A^2}} .
\end{eqnarray}
Combining with Abel's integral solution of equation (\ref{eq:sx}):
\begin{equation}
  \alpha_{\rm X}(T_{\rm e})[n_{\rm e}(r)]^2=
  \frac{1}{\pi A_y d_A}\int^{r/d_A}_{\infty}\frac{dS_x}{d\theta}
  \frac{d\theta}{\sqrt{\theta^2-(r/d_A)^2}},
\end{equation}
one can obtain the temperature profile $T_{\rm e}(r)$ by
\begin{equation}
  T_{\rm e}(r) = 
  {
  {\displaystyle \int_{\infty}^{r/d_A} 
  {d [S_x(\theta)T_{\rm X}(\theta)] \over d\theta}
  {d\theta \over \sqrt{\theta^2 - r^2/d_A^2}}} 
  \over
  {\displaystyle \int^{r/d_A}_{\infty}\frac{dS_x}{d\theta}
  \frac{d\theta}{\sqrt{\theta^2-(r/d_A)^2}}}
  }.
  \label{eq:tprof2} 
\end{equation}
Equation \ref{eq:tprof2} is more suitable for reconstructing the
temperature profile than equation (\ref{eq:tprof1}) in that all the
necessary information can be obtained from the X-ray observation alone,
and that it is independent of the specific form of the X-ray
emissivity $\alpha_{\rm X}(T_{\rm e})$. The use of equation
(\ref{eq:txproj}) requires sufficiently good spatial and energy
resolutions, both of which are feasible only with future X-ray
satellites including AXAF and XMM.

All the above procedures (eqs.[\ref{eq:nprof}], [\ref{eq:tprof1}] and
[\ref{eq:tprof2}]) involve the integral of the form:
\begin{equation}
  \label{eq:algorithm}
  \int^{r/d_A}_{\infty}\frac{df(\theta)}{d\theta}\frac{d\theta}
        {\sqrt{\theta^2-(r/d_A)^2}} .
\end{equation}
In what follows we numerically evaluate the integral as
\begin{equation}
  \label{eq:algorithm2}
  \sum_{i=i_{\rm min}}^{i_{\rm max}}
 \frac{f_i-f_{i+1}}{\sqrt{\theta_{i+\frac{1}{2}}^2-(r/d_A)^2}},
\end{equation}
where the observed quantity $f_i$ ($i=1,i_{\rm max}$) is given at the
discrete angular radius $\theta_i$, and $\theta_{i+\frac{1}{2}} =
(\theta_i+\theta_{i+1})/2$. $i_{\rm max}$ is the index for the
outermost bin and $i_{\rm min}$ indicates the bin which corresponds to
$r/d_A$. In practice we adopt $i_{\rm max}=30$ for definiteness. In fact
one could improve the above evaluation by {\it pre-smoothing} the
dataset to reduce the discrete noise component.  While such a
procedure is inevitably model-dependent and we did not attempt one in
what follows, this would be practically useful in dealing with the
actual observational data and improve the results below.

\vspace*{-0.5cm}

\section{Statistical and systematic errors in the reconstruction}

\subsection{models of gas density and temperature profiles}

Strictly speaking one needs to observe $S_x(\theta)$ and $S_y(\theta)$
for $r/d_A < \theta < \infty$ in order to determine $n_e(r)$ and
$T_e(r)$.  In practice, however, the observable fluxes are limited to
the finite extent of clusters, and also contaminated by the angular
resolution of the telescope and the detector noise especially at outer
regions. Moreover realistic clusters are not spherically symmetric to
some degree. In this section, we employ several cluster models with
different density and temperature profiles, and examine how the above
reconstruction method works in taking account of such observational
limitations.  More specifically, we examine the following four models:

\noindent {\bf Model A:} this is a conventional isothermal
$\beta$-model described by
\begin{equation} \label{eq:modela} n_e(r) = {n_{e0} \over
[1+(r/r_c)^2]^{3\beta/2} } , \qquad T_e(r) = T_{e0} . 
\end{equation}
We adopt a typical value for clusters, $\beta=2/3$, for definiteness. 
The values of the other parameters, $n_{e0}$, $T_{e0}$, and $r_c$, are
irrelevant for our current purpose since all the results can be
rescaled appropriately.

\noindent {\bf Model B:} consider a family of 
the dark matter halo profiles given by
\begin{equation}
\label{eq:haloprofile}
\rho_\dm(x) = {\delta_{\dm} \over x^\mu
(1+x^\nu)^\lambda },
\end{equation}
where $x \equiv r/r_s$ is the dimensionless radius in units of the
characteristic scale $r_s$, and $\delta_{\dm}$ is the amplitude of the
profile. NFW claimed that a halo profile with $(\mu, \nu,
\lambda)=(1,2,1)$ is the universal shape fairly independent of the
cosmological initial conditions (see also Fukushige \& Makino 1997 and
Moore et al. 1998).

Suto, Sasaki \& Makino (1998) found an analytical solution for the gas
density and temperature profiles when the gas obeys the polytropic
equation of state $T_{\rm e}(r) = T_{\rm e0}[n_e(r)/n_{e0}]^{1/n}$ and
its self-gravity can be neglected compared with the dark halo. Their
solution is written as
\begin{equation}
 n_{\rm e}(r) = n_{\rm e 0} [1-B_p f(x)]^n ,
\qquad
 T_{\rm e}(r) = T_{\rm e 0} [1-B_p f(x)] ,
\end{equation}
where
\begin{eqnarray}
\label{eq:fxdef}
f(x) &\equiv& \int_0^x {u^{1-\mu} \over (1+u^\nu)^\lambda} du 
- {1 \over x}\int_0^x {u^{2-\mu} \over (1+u^\nu)^\lambda} du ,
\end{eqnarray}
\begin{equation}
\label{eq:bp}
  B_{\rm p} \equiv \frac{4 \pi G}{n+1} 
\frac{\mu_{g}m_p \delta_\dm r_s^2 }{k T_{e0}} ,
\end{equation}
$G$ is the gravitational constant, $k$ is the Boltzmann constant,
$\mu_g$ is the mean molecular weight of the gas, and $m_p$ is the
proton mass.

Therefore given the dark matter halo profile, the gas density and
temperature profiles are specified by the additional two parameters
$n$ and $B_{\rm p}$. We consider the following two models (model B and
C) which have the analytic solutions for $f(x)$. In the NFW model
($\mu=1$, $\nu=2$, and $\lambda=1$),
\begin{equation}
f(x) = 1 - {\ln(1+x) \over x}.
\end{equation}
We adopt $n=12$, $B_p=1.0$ for model B so that the resulting
$S_x(\theta)$ has a fairly similar profile to that of model A
(corresponding to $\beta =2/3$).

\noindent {\bf Model C:} halo model with
$\mu=3/2$, $\nu=1$, and $\lambda=3/2$ has
\begin{equation}
f(x) = 2\sqrt{1+x \over x} -
{2 \over x} \ln\left( \sqrt{x} + \sqrt{1+x} \right) .
\end{equation}
We adopt $n=11$, $B_p=0.5$, again so that the resulting $S_x(\theta)$
resembles the profile of model A.

\noindent {\bf Model D:} we select a simulated cluster (cluster A at
$z=0$ in Yoshikawa, Itoh, \& Suto 1998) in order to examine a possible
effect of the non-sphericity. The cluster has lower temperature in the
central region unlike the above three models; in this sense also, it
can be regarded as a good example to test the robustness of the
present reconstruction method. In practice, we find that the
circularly averaged profiles of its X-ray and SZ surface brightness of
the cluster are well approximated by
\begin{equation} \label{eq:modele} S_x(\theta) = {S_x(0) \over
[1+(d_A\theta/r_X)^2]^{3\beta_X-1/2} } , \qquad S_y(\theta) = {S_y(0)
\over [1+(d_A\theta/r_y)^2]^{3\beta_y/2-1/2} } , 
\end{equation}
with $\beta_X=0.69$, $r_X=0.13$Mpc, $\beta_y=0.86$, and $r_y=0.18$Mpc.
Therefore we adopt the above spherical fits (with appropriate errors;
see the next subsection) as input data, and examine the extent to
which the reconstructed density and temperature profiles agree with
the spherically averaged ones directly computed from the simulated
data.

\subsection{error assignment and reconstructed profiles}

\begin{figure}[tbph]
\begin{center}
  \leavevmode
  \psfig{file=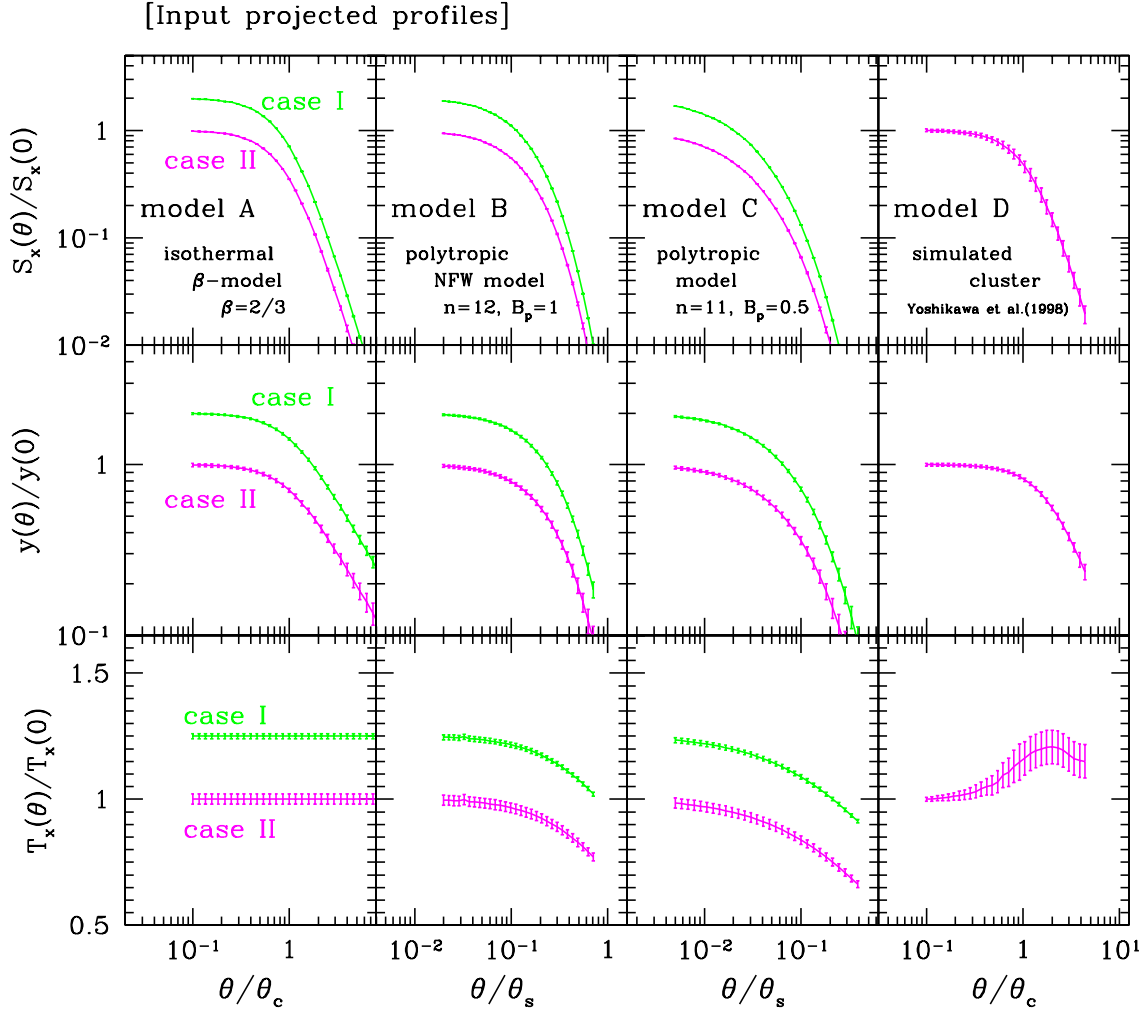,width=15cm}
\caption{Input profiles of scaled X-ray surface brightness (upper panels), 
  the SZ flux (middle panels) and emission-weighted temperature (lower
  panels) for models A to D. For models A, B, and C, the smaller and
  larger error bars correspond to our cases I and II of error
  assignment, respectively (eqs.  [\protect\ref{eq:case1}\protect] and
  [\protect\ref{eq:case2}\protect]).  The curves corresponding to case
  I are artificially shifted upward for an illustrative purpose.  For
  model D, the error bars indicate the $1\sigma$ deviation from
  spherical symmetry obtained from the numerical simulation (Yoshikawa
  et al. 1998). The angular core radius $\theta_c$ in model D is
  computed from the core radius $r_c$ of the $\beta$-model fit to the
  spherically averaged density profile of the simulated cluster. Solid
  lines indicate the true profile {\it before} assigning the error.}
\label{fig:inputprof}
\end{center}
\end{figure}

Figure~\ref{fig:inputprof} shows the input projected profiles of
bolometric X-ray flux, y-parameter and X-ray emission-weighted
temperature (eqs. [\ref{eq:sx}], [\ref{eq:sy}], and [\ref{eq:txproj}])
for models A to D described in \S 3.1.  In order to estimate
statistical and possible systematic errors to our method due to
observational uncertainties and deviation from spherical symmetry, we
perform the reconstruction after adding the following Gaussian
distributed errors. More specifically we examine models A, B, and C to
estimate errors from observational uncertainties, and model D from the
deviation from spherical symmetry.

The relevance of the error assignment should sensitively depend on
both the observing facilities and the specific target clusters one has
in mind.  For models A to C, we consider two cases (I and II) for
error assignment; for $S_x$ the assumed errors are fairly realistic
even with the current X-ray satellites (e.g., Briel \& Henry 1996) at
least for relatively near and bright clusters, but for $S_y$ and
$T_{\rm X}$ the errors should be feasible only with the future
interferometer facilities including Japanese LMSA (Large Millimeter
and Submillimeter Array; Kawabe, private communication), and future
X-ray satellites with high sensitivity such as XMM with sufficiently
large integration time. For a given error of $S_x$, one may roughly
scale the resulting error-bars in Figure~\ref{fig:reconstprof} for the
different level of the errors in $S_y$ and $T_{\rm X}$. With this in
mind, we consider two specific cases; for the case I, we assign the
$1\sigma$ errors as
\begin{equation}
  \Delta S_x(\theta) = 10^{-4} S_x(0), \hspace{0.5cm}
  \Delta S_y(\theta) = 10^{-2} S_y(0), \hspace{0.5cm}
  \Delta T_{\rm X}(\theta) = 10^{-2} T_{\rm X}(\theta)
\label{eq:case1}  
\end{equation}
at each $\theta$, and for the case II they are
\begin{equation}
  \Delta S_x(\theta) = 10^{-3} S_x(0), \hspace{0.5cm}
  \Delta S_y(\theta) = 2\times10^{-2} S_y(0), \hspace{0.5cm}
  \Delta T_{\rm X}(\theta) = 5\times10^{-2} T_{\rm X}(\theta) .
\label{eq:case2}  
\end{equation}
Error bars in Figure \ref{fig:inputprof} indicate those $1\sigma$
dispersion of Gaussian errors in the case of I and II (all the curves
for case I are artificially shifted upward so that the two cases are
easily distinguished). For model D, on the other hand, we assign the
$1\sigma$ dispersion of the fluxes around the circularly averaged
values directly computed from simulation data profile by Yoshikawa,
Itoh, \& Suto (1998), and do not attempt to include the observational
errors. We sample the data using 30 radial bins in logarithmically
equal intervals.

According to the above procedure, we construct 200 realizations for
each model with different random number for the error assignment, and
then perform the reconstruction.  Figure~\ref{fig:reconstprof} shows
the reconstructed profiles of gas density (top panels), the ratio
between them (upper middle panels), temperature (lower middle and
bottom panels) for each model.  The lower middle panels show the
temperature profiles reconstructed from X-ray surface brightness and
the SZ flux (eq. [\ref{eq:tprof1}]) and the bottom panels show that
reconstructed from X-ray surface brightness and emission-weighted
temperature (eq.[\ref{eq:tprof2}]). Solid lines correspond to the {\it
  true} profiles for models A to C (again the results for case I are
artificially shifted upward for an illustrative purpose), and the
spherically averaged profiles for model D.

The reconstructed profiles are plotted with their $1\sigma$ error-bars
computed from the 200 realizations. Since the main contribution for
the projected fluxes comes from the radius around $r_c$, the estimates
become less reliable either for $r \ll r_c$ or $r \gg r_c$.  Also the
reconstruction works better for $n_e(r)$ than for $T_e(r)$ since the
X-ray emissivity is more sensitive to the former ($\propto n_e^2
T_e^{1/2}$ for the thermal bremsstrahlung). The upper and lower solid
lines in model D represent the $\pm 1\sigma$ deviation from the
spherically averaged profile, which illustrate the degree of
non-sphericity of the simulated cluster.  The degree of asphericity in
our simulated cluster is comparable to, or even larger than, the error
due to the reconstruction procedure.  Since our simulated cluster
seems to be typical in the light of the distribution of non-sphericity
in the observed cluster sample (Mohr et al. 1995), the non-sphericity
of that degree does not seriously degrade the current methodology
although the inclusion of non-spherical effect is definitely an
important next step (Zaroubi et al. 1998).

\section{Discussion and conclusions}

We have presented a method to reconstruct the radial profiles of gas
density and temperature in clusters of galaxies.  Since most existing
techniques start with some sort of empirical cluster gas profile and
attempt to find the best-fit in that modeling, the result is fairly
model-dependent.  Our current method is in marked contrast to the
previous ones in that the radial profiles can be reconstructed in a
non-parametric manner without assuming the equation of state or
hydrostatic equilibrium of cluster gas if high-resolution projected
profiles of X-ray and SZ observations are given.

\begin{figure}[tbph]
\begin{center}
  \leavevmode
  \psfig{file=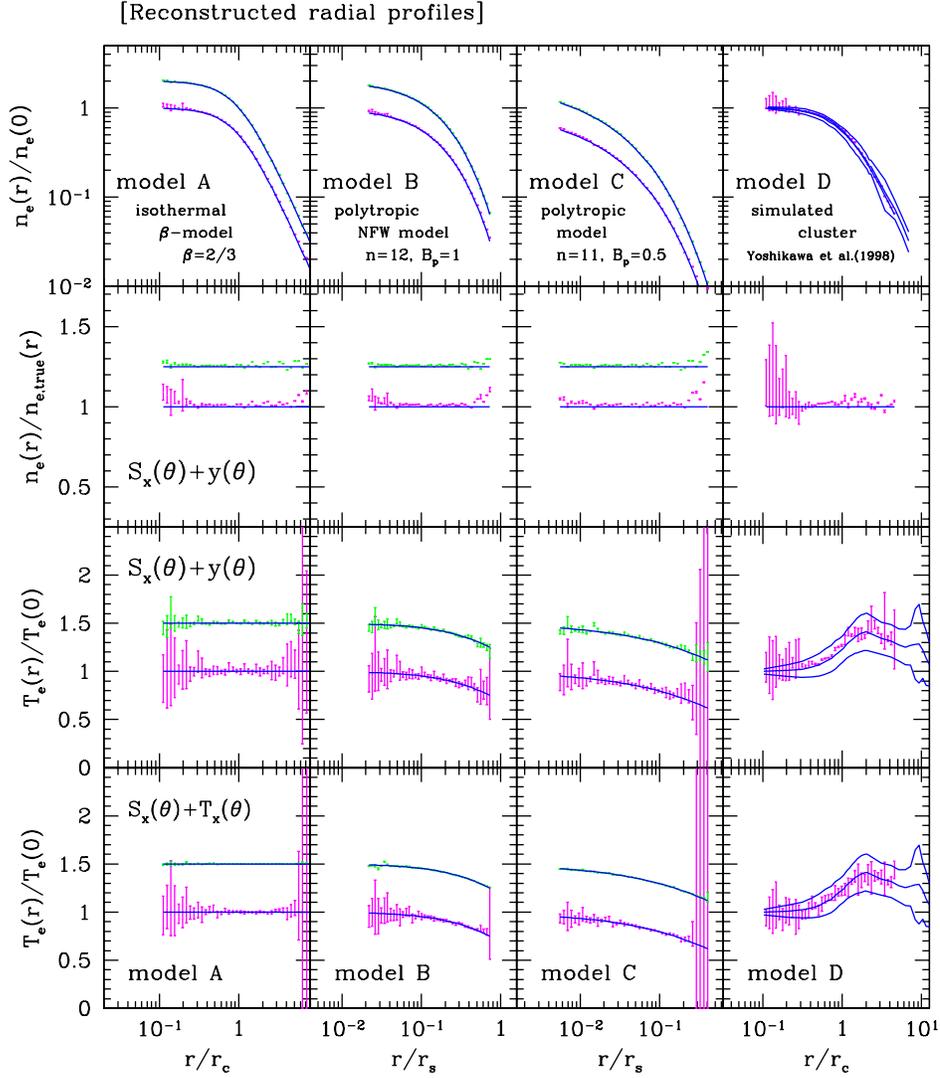,width=15cm}
\caption{Reconstructed gas density and temperature profiles
  for models A to D. {\it Top panels:} reconstructed $n_e(r)$ in units
  of the central value; {\it Upper middle panels:} reconstructed
  $n_e(r)$ in units of the true density $n_{e,true}(r)$; {\it Lower
    middle panels:} $T_e(r)$ reconstructed with X-ray and SZ fluxes in
  units of the central value; {\it Bottom panels:} $T_e(r)$
  reconstructed with X-ray flux and emission-weighted projected
  temperature in units of the central value.  For models A to C, solid
  lines indicate the true model profile, and the error bars correspond
  to those shown in Figure 1. The curves corresponding to case I are
  artificially shifted upward for an illustrative purpose.  For model
  D, the middle solid lines in each panel represent the spherically
  averaged profiles, and the upper and lower solid lines indicate the
  $\pm 1\sigma$ dispersion of the simulated cluster around the
  spherically averaged profile.}
\label{fig:reconstprof}
\end{center}
\end{figure}

We apply this method to three analytic spherical models (A to C) with
gas density and temperature profiles and one non-spherical simulated
cluster. For models A to C, the fractional uncertainties of
reconstructed gas density profiles are less than 1\% in our case I
while they amount to 20--30\% at central regions and 2--3\% at outer
regions in our case II.  On the other hand, the estimate of
temperature is sensitive to the accuracy of the SZ flux or X-ray
emission weighted temperature as in Figure~\ref{fig:reconstprof}. With
the degree of asphericity in our simulated cluster, the resulting
systematic error is smaller than our adopted statistical errors.  In
fact while our error assignment for X-ray temperature and the SZ
fluxes may be rather optimistic, one can rescale the resulting error
bars according to the real observation if necessary.

Although we have presented the analysis on the basis of the spherical
symmetry, the same methodology is in principle applicable to other
non-spherical systems if one accepts some additional assumption on the
density distribution along the line of sight. In this context, one of
the most practically important ways is to assume a bilateral symmetry
(e.g., Fabricant, Rybicki \& Gorenstein 1984; Hughes \& Birkinshaw
1998). In this case, all the results presented in this paper is
readily applicable simply by an appropriate choice of the coordinate
transformation.

Considering the future extensive observation of clusters of galaxies
in X-ray and radio bands with upcoming facilities such as ABRIXAS,
AXAF, XMM, and LMSA, high-resolution images of clusters of galaxies in
X-ray and radio bands will be available with unprecedented high
signal-to-noise ratio and the reconstruction of profiles of clusters
of galaxies with sufficient accuracy should be feasible. More
quantitative discussion of the error and the resulting accuracy of the
reconstruction should definitely depend on the specific target
cluster, and we are not able to proceed further at this moment.
Nevertheless we believe that the current methodology becomes useful in
near future. Furthermore with such a reliable reconstructed radial
profiles of clusters of galaxies, even if feasible only for relatively
near and bright ones, one can revisit various cosmological issues, in
which simplified and/or empirical models have been adopted, including
the estimates of Hubble constant (Kobayashi, Sasaki \& Suto 1996),
peculiar velocity of clusters of galaxies (Yoshikawa, Itoh, \& Suto
1998), the $L-T$ relation which is inconsistent with observations if
derived from a simple scaling argument, and physics cooling flow.

\bigskip

We thank Pat Henry for information of the current observational errors
of the X-ray fluxes, and Ryohei Kawabe for useful discussion on the
current and future observing facilities in radio, millimeter and
submillimeter bands and their expected performance.  We also thank an
anonymous referee for several comments which helped improve the
earlier version of the paper.  Numerical computations were carried out
on VPP300/16R and VX/4R at ADAC (the Astronomical Data Analysis
Center) of the National Astronomical Observatory, Japan, as well as at
RESCEU (Research Center for the Early Universe, University of Tokyo)
and KEK (National Laboratory for High Energy Physics, Japan). This
research was supported in part by the Grants-in-Aid by the Ministry of
Education, Science, Sports and Culture of Japan (07CE2002) to RESCEU,
and by the Supercomputer Project (No.97-22) of High Energy Accelerator
Research Organization (KEK).

\newpage
\bigskip 

\baselineskip13pt
\parskip2pt
\newpage
\centerline{\bf REFERENCES}
\bigskip

\def\apjpap#1;#2;#3;#4; {\pp#1, {#2}, {#3}, #4}
\def\apjbook#1;#2;#3;#4; {\pp#1, {#2} (#3: #4)}
\def\apjppt#1;#2; {\pp#1, #2.}
\def\apjproc#1;#2;#3;#4;#5;#6; {\pp#1, {#2} #3, (#4: #5), #6}

\apjppt Birkinshaw, M. 1998;Phys.Rep., in press (astro-ph/9808050);
\apjpap Birkinshaw, M., Hughes, J.P. \& Arnaud, K.A. 1991;ApJ;379;466;
\apjpap Briel, U.G., \& Henry, J. P. 1996;ApJ;472;131; 
\apjpap Bryan, G.L., \& Norman, M. 1998;ApJ;495;80;
\apjpap Cooray, A.R., Grego, L., Holzapfel, W.L., Joy, M., \&
Carlstrom, J.E. 1998;AJ;115;1388;
\apjpap Couchman, H.M.P., Thomas, P.A. \& Pearce, F.R. 1995;
ApJ;452;797;
\apjpap Eke, V. R., Cole, S., \& Frenk, C. S. 1996;MNRAS;282;263;
\apjpap Eke, V.R, Navarro, J.F., \& Frenk, C.S. 1998;ApJ;503;569;
\apjpap Evrard, A. E., \& Henry, J. P. 1991;ApJ;383;95; 
\apjpap Fabricant, D., Rybicki, G., \& Gorenstein, P. 1984;ApJ;286;186;
\apjpap Fukushige, T., \& Makino, J. 1997;ApJ;477;L9;
\apjpap Henry, J. P., \& Arnaud, K. A. 1991;ApJ;372;410; 
\apjpap Hughes, J.P., \& Birkinshaw, M. 1998;ApJ;501;1; 
\apjpap Inagaki, Y., Suginohara, T.,\& Suto, Y. 1995;PASJ;47;411;
\apjpap Kitayama, T. \& Suto, Y. 1997;ApJ;490;557;
\apjpap Kitayama, T., Sasaki,S., \& Suto, Y. 1998;PASJ;50;1;
\apjpap Kobayashi, S., Sasaki,S., \& Suto, Y. 1996;PASJ;48;L107;
\apjppt Komatsu, E., Kitayama, T., Suto, Y., Hattori, M., 
  Kawabe, R., Matsuo, H., Schindler, S., 
   \& Yoshikawa, K. 1998;Nature, submitted;
\apjpap Makino, N., Sasaki, S., \& Suto, Y. 1998;ApJ;497;555;
\apjpap Mohr, J.J., Evrard, A.E., Fabricant, D.G., \& Geller, M.J. 1995;ApJ;447;8;
\apjpap Moore, B., Governato, F., Quinn, T., Stadel, J.,\& Lake, G. 1998;
   ApJ;499;L5;
\apjpap Navarro, J.F., Frenk, C.S., \& White, S.D.M. 1997;ApJ;490
   ;493 (NFW);
\apjpap Rephaeli, Y. 1995;ARA\&A;33;541;
\apjpap Shimasaku, K. 1998;ApJ;489;501; 
\apjpap Silk, J. \& White, S.D.M. 1978; ApJ;226;L103;
\apjpap Sunyaev R.A. \& Zel'dovich Ya.B.  1972;Comments on 
   Astrophys.\& Space Phys.;4;173;
\apjpap Sunyaev R.A. \& Zel'dovich Ya.B.  1980;MNRAS;190;413;
\apjpap Suto, Y., Sasaki, S., \& Makino, N. 1998;ApJ;509;
December 20 issue, in press;
\apjppt Tegmark, M., Eisenstein, D.J.,  Hu, W., \& Kron, R. 1998;
ApJ, in press (astro-ph/9805117);
\apjpap Viana, P. T. P., \& Liddle, A. R. 1996;MNRAS;281;323;
\apjpap White, S. D. M., Efstathiou, G., \& Frenk, C. S. 1993;MNRAS;262;1023;
\apjpap White, S. D. M., Navarro, J. F., Evrard, A. E., 
   \& Frenk, C. S. 1993;Nature;366;429;  
\apjpap Wu, X.P., \& Fang, L.Z. 1997;ApJ;483;62; 
\apjpap Yoshikawa, K., Itoh, M., \& Suto, Y. 1998;PASJ;50;203;
\apjpap Zaroubi, S., Squires, G., Hoffman, Y., Silk, J. 1998;ApJ;500;L87;

\newpage

\bigskip
\bigskip

\end{document}